\documentclass[journal]{IEEEtran}
\usepackage{cite}
\usepackage{amsmath,amssymb,amsfonts}
\usepackage{algorithm}
\usepackage{algpseudocode}
\usepackage{graphicx}
\usepackage{url}
\usepackage{booktabs}
\usepackage{textcomp}
\usepackage{amsthm}
\usepackage{stmaryrd,txfonts}
\usepackage{mathrsfs}
\usepackage{tikz}
\usetikzlibrary{intersections,arrows.meta}
\usepackage{microtype}
\usepackage{float}
\usepackage{xcolor}
\usepackage[colorlinks=true,linkcolor=blue,citecolor=blue,urlcolor=blue]{hyperref}

\makeatletter
\def\@cite#1#2{\textcolor{blue}{[{#1\if@tempswa , #2\fi}]}}
\makeatother
\let\oldeqref\eqref
\renewcommand{\eqref}[1]{\textcolor{blue}{\oldeqref{#1}}}

\newtheorem{definition}{Definition}

\makeatletter
\newenvironment{breakablealgorithm}
{
        \begin{center}
            \refstepcounter{algorithm}
            \hrule height.8pt depth0pt \kern2pt
            \renewcommand{\caption}[2][\relax]{
                {\raggedright\textbf{\ALG@name~\thealgorithm} ##2\par}%
                \ifx\relax##1\relax
                \addcontentsline{loa}{algorithm}{\protect\numberline{\thealgorithm}##2}%
                \else
                \addcontentsline{loa}{algorithm}{\protect\numberline{\thealgorithm}##1}%
                \fi
                \kern2pt\hrule\kern2pt
            }
        }{
        \kern2pt\hrule\relax
    \end{center}
}
\makeatother

\begin{document}

\title{Dynamic Layered Decoding Scheduling for LDPC Codes Aided by Check Node Unsatisfied Probabilities}
\author{Chenyuan Jia,
        Dongxu Chang,
        Ruiyuan Wang,
        Guanghui Wang,
        Guiying Yan,
        Cunquan Qu
        \thanks{This work is partially supported by the National Key R\&D Program of China, (2023YFA1009600) (Corresponding author: Guanghui Wang).}
        \thanks{C. Jia, D. Chang, G. Wang and C. Qu are with the School of Mathematics, Shandong University, Shandong, China (e-mail: chenyuanjia@mail.sdu.edu.cn; dongxuchangmail@163.com; ghwang@sdu.edu.cn; cqqu@sdu.edu.cn).}
        \thanks{G. Yan is with the Academy of Mathematics and Systems Science, CAS, University of Chinese Academy of Sciences, Beijing, 100190 China (e-mail: yangy@amss.ac.cn).}
        \thanks{R. Wang is with the Tsinghua Shenzhen International Graduate School, Tsinghua University, Shenzhen, China (e-mail: ry-wang24@mails.tsinghua.edu.cn).}
        }

\maketitle

\begin{abstract}
This letter revisits update ordering in layered belief propagation (LBP) decoding of low-density parity-check (LDPC) codes. The closest probability-based schedule orders layers by check node unsatisfied probabilities estimated only from the channel observations, although these probabilities change once decoding messages are exchanged. We therefore refresh the check node unsatisfied probabilities during decoding and use them as dynamic priorities. The first schedule, Dyn-EBP, selects the most reliable available check node while ensuring that every check node is updated once in each iteration. The second schedule, Dyn-PEBP, adds a linear update-count penalty and allows limited repeated updates without letting a small subset of check nodes dominate the schedule. For 5G new radio LDPC base graph 1 codes with five iterations, the proposed schedules yield small BLER reductions relative to the channel-only probability schedule and remain competitive with LBP, LPHD scheduling, and RD-RBP. The gain is modest, but it shows that probability-based scheduling benefits from message-level refinement.
\end{abstract}


\section{Introduction}\label{sec:introduction}
\IEEEPARstart{L}{ow-density} parity-check (LDPC) codes \cite{gallager1962low} are standard components in modern communication systems, including 5G new radio (NR) \cite{3gpp20185g} and IEEE 802.11 wireless networks \cite{ieee2007ieee}. Since the near-capacity behavior reported by MacKay and Neal \cite{mackay1997near}, belief propagation (BP) decoding \cite{mackay1999good} has remained one of the main practical decoders for LDPC codes because it combines sparse-graph structure, low implementation cost, and strong empirical performance.

The conventional flooding schedule updates all variable-to-check (V2C) messages and all check-to-variable (C2V) messages in separate global stages. This parallelism is attractive in hardware, but new information generated in one stage cannot be exploited until the next stage. Layered belief propagation (LBP) \cite{hocevar2004reduced} reduces this delay by updating the graph layer by layer, so that a newly computed message can influence later updates within the same iteration. This reuse of fresh information allows LBP to approach the performance of flooding BP with fewer iterations.

A remaining issue is that LBP does not prescribe which check node or layer should be updated first. The order can affect convergence, latency, and the usefulness of the messages passed later in the same iteration \cite{chang2024optimization}. Dynamic schedules address this issue by choosing the next update from quantities observed during decoding. Residual-based methods \cite{casado2007informed,gong2011effective,liu2014variable,Huilian2019Residual}, for example, promote nodes whose messages have changed sharply. Such criteria are effective in many cases, but they may also concentrate updates on a small part of the graph and leave other constraints underused.

The schedule in \cite{chang2025} takes a different view. It shows that, over both binary erasure and binary-input additive white Gaussian noise channels, an efficient layered schedule should update check nodes with lower unsatisfied probabilities earlier. The resulting design estimates these probabilities from channel information and uses the estimates to form a simple scheduling sequence. This criterion is well aligned with the role of a parity-check equation: a check node that is less likely to be unsatisfied is also more likely to send reliable extrinsic information to its neighbors.

The present letter refines that idea by updating the check node unsatisfied probabilities as decoding proceeds. Two schedules are proposed. Dyn-EBP enforces one update of every check node in each iteration and refreshes only the priorities affected by the latest C2V messages. Dyn-PEBP introduces a penalty on the update count, allowing repeated updates when the local evidence is strong while discouraging the redundant consecutive updates that can occur across iteration boundaries. Simulations on 5G NR LDPC codes show small but repeatable BLER gains over the channel-only schedule of \cite{chang2025}. The resulting gain is incremental; it indicates that the probability criterion becomes more accurate when it is refreshed during message passing.

The remainder of the letter is organized as follows. Section~\ref{pre} reviews BP, LBP, and scheduling sequences. Section~\ref{s3} gives the check node unsatisfied probability used as the scheduling metric. Section~\ref{s4} presents Dyn-EBP, Dyn-PEBP, and their complexity. Section~\ref{sim} reports numerical results, and Section~\ref{con} concludes the letter.

\section{Preliminaries}\label{pre}

\subsection{Belief Propagation for LDPC Codes}
An LDPC code can be represented by a Tanner graph with $N$ variable nodes $v_j$, $j=1,2,\ldots,N$, and $M$ check nodes $c_i$, $i=1,2,\ldots,M$. Let $\mathcal{N}(v)$ denote the neighbors of node $v$. In the LLR domain, the V2C update in the $l$-th iteration is
\begin{equation}
    m_{v_j \rightarrow c_i}^{(l)}=C_{v_j}+\sum_{k \in \mathcal{N}(v_j)\setminus\{c_i\}} m_{c_k \rightarrow v_j}^{(l-1)},
    \label{v2c}
\end{equation}
and the C2V update is
\begin{equation}
    m_{c_i \rightarrow v_j}^{(l)}=2\tanh^{-1}\!\left[\prod_{h \in \mathcal{N}(c_i)\setminus\{v_j\}} \tanh\left(m_{v_h \rightarrow c_i}^{(l-1)}/2\right)\right],
    \label{c2v}
\end{equation}
where $C_{v_j}$ is the channel LLR of $v_j$. Initial messages are set to zero. Decoding stops when all parity checks are satisfied or when the maximum number of iterations is reached. The a posteriori LLR of variable node $v_j$ is
\begin{equation}
    L_j^{(l)}=C_{v_j}+\sum_{i\in \mathcal{N}(v_j)}m_{c_i\rightarrow v_j}^{(l-1)},
    \label{appllr}
\end{equation}
and the hard decision is
\begin{equation}\label{hd}
    x_j=\begin{cases}
    0, & L_j\geq 0,\\
    1, & L_j<0.
    \end{cases}
\end{equation}

\subsection{Layered Belief Propagation and Scheduling Sequences}
In LBP \cite{hocevar2004reduced}, the decoder selects a check node, updates the V2C messages on all incident edges, and then produces the corresponding C2V messages. The updated messages immediately modify the LLRs of adjacent variable nodes and can be used by later check node updates in the same iteration.

\begin{definition}[scheduling sequence]
A scheduling sequence is the ordered list of check node indices selected by the LBP decoder. When the same order is used in every iteration, the sequence is specified by the check node order within one iteration.
\end{definition}

The same parity-check matrix can therefore support many layered schedules. A good order should propagate reliable information early, avoid unnecessary repetitions, and keep the graph sufficiently covered within each decoding round.

\section{Check Node Unsatisfied Probability}\label{s3}
A check node is unsatisfied when its parity-check equation is violated. Equivalently, an odd number of its neighboring variable nodes are decoded incorrectly. Given the current LLR $L_j$, the flip probability of variable node $v_j$ is
\begin{equation}\label{pv}
    p_j^{v_\epsilon}=\frac{1}{1+\left[\lambda^{-1}I(L_j<0)+\lambda I(L_j\geq0)\right]e^{|L_j|}},
\end{equation}
where $\lambda=P(x_j=0)/P(x_j=1)$ and $I(\cdot)$ is the indicator function. For the usual equiprobable input case, $\lambda=1$, and \eqref{pv} becomes \cite{Land2000LoglikelihoodVA,6325232}
\begin{equation}
    p_j^{v_\epsilon}=\frac{1}{1+e^{|L_j|}}.
    \label{varprob}
\end{equation}
Under the standard extrinsic-message independence approximation, the probability that a parity check contains an odd number of erroneous neighboring variable nodes follows Gallager's parity identity \cite{gallager1962low}. Thus the unsatisfied probability of check node $c_i$ is
\begin{equation}\label{cv}
    p_i^{c_{\mathrm{u}}}=\frac{1}{2}\left[1-\prod_{j\in\mathcal{N}(c_i)}\left(1-\frac{2}{1+e^{|L_j|}}\right)\right],
\end{equation}
where $i=1,2,\ldots,M$. This expression is the scheduling priority used below. The derivation is direct: the parity check fails exactly when the number of neighboring variable node flips is odd, and the odd-parity probability of independent Bernoulli events is given by the product in \eqref{cv}.

\section{Proposed Methods}\label{s4}

\subsection{Dyn-EBP Decoding}\label{B}
The channel-only schedule in \cite{chang2025} fixes the priority of a check node before message passing starts. During LBP decoding, however, each C2V update changes the neighboring variable node LLRs in \eqref{appllr}; through \eqref{cv}, it also changes the unsatisfied probabilities of adjacent check nodes. Dyn-EBP uses this local change to update the schedule on the fly.

Initially, $m_{v_j\rightarrow c_i}$ is set to $C_{v_j}$ and the unsatisfied probabilities $p_i^{c_{\mathrm{u}}}$ are computed for all check nodes. The decoder stores the check nodes in a priority set $P$ sorted in ascending order of $p_i^{c_{\mathrm{u}}}$. At each step, Dyn-EBP selects the first element of $P$, updates all messages attached to that check node, refreshes the affected neighboring probabilities, and reorders $P$. The selected check node is removed from $P$ until the next iteration, which guarantees that every check node is updated once per iteration.

\begin{figure}[!t]
    \centering
    \begin{tikzpicture}[>=Stealth,scale=0.86]
        \draw (-2,0) circle[radius=0.35];
        \filldraw[fill=blue!20!white,draw=blue!50!black] (-1,0) circle[radius=0.35];
        \filldraw[fill=blue!20!white,draw=blue!50!black] (0,0) circle[radius=0.35];
        \filldraw[fill=blue!20!white,draw=blue!50!black] (1,0) circle[radius=0.35];
        \draw (2,0) circle[radius=0.35];
        \draw (3,0) circle[radius=0.35];
        \draw (4,0) circle[radius=0.35];
        \draw (-2,-0.35) -- (-1.01,-2);
        \draw (-1,-0.35) -- (-1.01,-2);
        \draw (-1,-0.35) -- (-0.335,-1.175);
        \draw[<-] (-0.335,-1.175) -- (0.33,-2);
        \draw (0,-0.35) -- (0.165,-1.175);
        \draw[<-] (0.165,-1.175) -- (0.33,-2);
        \draw (1,-0.35) -- (0.665,-1.175);
        \draw[<-] (0.665,-1.175) -- (0.33,-2);
        \draw (1,-0.35) -- (1.67,-2);
        \draw (2,-0.35) -- (1.67,-2);
        \draw (2,-0.35) -- (3.01,-2);
        \draw (3,-0.35) -- (3.01,-2);
        \draw (4,-0.35) -- (3.01,-2);
        \filldraw[fill=yellow!20!white,draw=yellow!50!black] (-1.26,-2.5) rectangle (-0.76,-2);
        \filldraw[fill=red!20!white,draw=red!50!black] (0.08,-2.5) rectangle (0.58,-2);
        \filldraw[fill=yellow!20!white,draw=yellow!50!black] (1.42,-2.5) rectangle (1.92,-2);
        \draw (2.76,-2.5) rectangle (3.26,-2);
    \end{tikzpicture}
    \caption{Local priority update in Dyn-EBP. After the red check node $c_k$ sends C2V messages to the blue variable nodes, only the neighboring yellow check nodes need their unsatisfied probabilities refreshed.}
    \label{exc}
\end{figure}

Fig.~\ref{exc} illustrates the local nature of the refresh. If $c_k$ is selected and sends $m_{c_k\rightarrow v_a}$, then only the check nodes in $\mathcal{N}(v_a)\setminus\{c_k\}$ can have changed priorities. This locality avoids recomputing the metric for every check node after every message update. Algorithm~\ref{debp} gives the resulting decoding procedure.

\begin{figure}[!t]
    \centering
    \begin{tikzpicture}[scale=0.80]
        \draw (0,0) rectangle (1,1) node[pos=.5] {$c_3$};
        \draw (1,0) rectangle (2,1) node[pos=.5] {$c_1$};
        \draw (2,0) rectangle (3,1) node[pos=.5] {$c_4$};
        \draw (3,0) rectangle (4,1) node[pos=.5] {$c_2$};
        \filldraw[fill=red!20!white,draw=red!50!black] (4,0) rectangle (5,1) node[pos=.5] {$c_5$};
        \node[above] at (-1.5,0.25) {iteration $=l$:};
        \filldraw[fill=red!20!white,draw=red!50!black] (0,-2) rectangle (1,-1) node[pos=.5] {$c_5$};
        \draw (1,-2) rectangle (2,-1) node[pos=.5] {$c_2$};
        \draw (2,-2) rectangle (3,-1) node[pos=.5] {$c_3$};
        \draw (3,-2) rectangle (4,-1) node[pos=.5] {$c_4$};
        \draw (4,-2) rectangle (5,-1) node[pos=.5] {$c_1$};
        \node[above] at (-1.7,-1.75) {iteration $=l+1$:};
    \end{tikzpicture}
    \caption{Successive repeated update at an iteration boundary. The check node $c_5$ is updated at the end of iteration $l$ and again at the beginning of iteration $l+1$.}
    \label{exn}
\end{figure}

A purely greedy dynamic schedule may produce the boundary effect shown in Fig.~\ref{exn}: the last check node in one iteration may also be the first in the next. Such an update uses almost the same incoming information twice and is therefore less useful than an update that first receives fresh messages from the surrounding graph. Dyn-EBP prevents this behavior within each iteration by enforcing full coverage before any check node is selected again.

\newpage

\begin{breakablealgorithm}\label{debp}
\caption{Dyn-EBP decoding for LDPC codes}
\begin{algorithmic}[1]
\State Initialize $m_{c_i\rightarrow v_j}=0$, $m_{v_j\rightarrow c_i}=C_{v_j}$, and $L_j=C_{v_j}$
\State Compute $p_i^{c_{\mathrm{u}}}$ for all check nodes and generate $P$
\While{$P\neq\emptyset$}
    \State Select the first element $p_k^{c_{\mathrm{u}}}$ in $P$
    \State Remove $p_k^{c_{\mathrm{u}}}$ from $P$
    \For{each $a\in\mathcal{N}(c_k)$}
        \State Generate and propagate $m_{v_a\rightarrow c_k}$
    \EndFor
    \For{each $a\in\mathcal{N}(c_k)$}
        \State Generate and propagate $m_{c_k\rightarrow v_a}$
        \State Update $L_a=L_a+m_{c_k\rightarrow v_a}$
        \For{each $b\in\mathcal{N}(v_a)\setminus\{c_k\}$}
            \State Recompute $p_b^{c_{\mathrm{u}}}$
        \EndFor
    \EndFor
    \State Reorder $P$
\EndWhile
\If{the stopping rule is not satisfied}
    \State Start the next iteration
\EndIf
\end{algorithmic}
\end{breakablealgorithm}

\subsection{Dyn-PEBP Decoding}
Dyn-PEBP relaxes the strict once-per-iteration rule. It retains the unsatisfied probability as the main priority but adds a penalty that increases with the number of times a check node has already been selected. The penalized priority is
\begin{equation}\label{pp}
    \tilde{p}_i^{c_{\mathrm{u}}}=p_i^{c_{\mathrm{u}}}+f(l_i),
\end{equation}
where $l_i$ is the update count of check node $c_i$. In the simulations, a linear penalty is used,
\begin{equation}
    f(l_i)=\gamma l_i,\qquad 0\leq \gamma\leq 1.
\end{equation}
Small $\gamma$ values make the schedule closer to a greedy probability minimizer, whereas large $\gamma$ values penalize repeated updates more strongly. Since $p_i^{c_{\mathrm{u}}}\in[0,1]$, the case $\gamma=1$ effectively recovers the coverage behavior of Dyn-EBP. The practical role of $\gamma$ is therefore to control how much repetition is allowed before all check nodes have been visited.

\subsection{Complexity Analysis}\label{comp}
Let $E$ be the number of edges in the Tanner graph, with $E=\bar d_v N=\bar d_c M$, where $\bar d_v$ and $\bar d_c$ are the average variable and check node degrees. The V2C and C2V message updates each require $O(E)$ operations per iteration. The additional cost comes from refreshing \eqref{cv}. A C2V message changes the LLR of one variable node; this affects the priorities of its neighboring check nodes, and each affected priority contains a product over the corresponding check node neighbors. The dominant probability-refresh cost is therefore $O(E\bar d_v\bar d_c)$ per iteration.

The dynamic ordering can be implemented with a priority queue. Updating every priority after every check node update would be unnecessarily expensive, requiring $O(M^2\log M)$ operations in a direct implementation. Instead, a Fibonacci heap \cite{1987Fibonacci} initializes the $M$ priorities in $\Theta(M)$ time and updates only the locally affected entries. Under the priority-update direction used in the schedule, each affected update costs at most $O(\bar d_c\bar d_v)$, and the reordering overhead per iteration is $O(M\bar d_c\bar d_v)=O(E\bar d_v)$. The total per-iteration complexity of both Dyn-EBP and Dyn-PEBP is thus $O(E\bar d_v\bar d_c)$. The same BP message equations are used as in ordinary LBP; the proposed methods change only the order in which check nodes are selected.

\begin{breakablealgorithm}\label{pebp}
	\caption{Dyn-PEBP decoding for LDPC codes}
	\begin{algorithmic}[1]
		\State Initialize $m_{c_i\rightarrow v_j}=0$, $m_{v_j\rightarrow c_i}=C_{v_j}$, $L_j=C_{v_j}$, and $l_i=0$
		\State Compute $\tilde{p}_i^{c_{\mathrm{u}}}$ for all check nodes and generate $P$
		\For{$t=1,2,\ldots,M$}
		\State Select the first element $\tilde{p}_k^{c_{\mathrm{u}}}$ in $P$
		\For{each $a\in\mathcal{N}(c_k)$}
		\State Generate and propagate $m_{v_a\rightarrow c_k}$
		\EndFor
		\For{each $a\in\mathcal{N}(c_k)$}
		\State Generate and propagate $m_{c_k\rightarrow v_a}$
		\State Update $L_a=L_a+m_{c_k\rightarrow v_a}$
		\For{each $b\in\mathcal{N}(v_a)$}
		\State Recompute $\tilde{p}_b^{c_{\mathrm{u}}}$
		\EndFor
		\EndFor
		\State Set $l_k=l_k+1$ and recompute $\tilde{p}_k^{c_{\mathrm{u}}}$
		\State Reorder $P$
		\EndFor
		\If{the stopping rule is not satisfied}
		\State Start the next iteration
		\EndIf
	\end{algorithmic}
\end{breakablealgorithm}

\section{Simulation Results}\label{sim}
The schedules are evaluated on 5G NR LDPC BG1 codes across blocklengths, rates, and signal-to-noise ratios. All decoders use five iterations. The baselines are LBP, least-punctured and highest-degree (LPHD) scheduling \cite{wang2020two}, RD-RBP \cite{Huilian2019Residual}, and the channel-only probability schedule of \cite{chang2025}. For quasi-cyclic layer scheduling, each layer's priority is the average unsatisfied probability of its check nodes in the lifted graph. Puncturing sets the initial unsatisfied probabilities of adjacent check nodes to 0.5; consequently, the first updated layer corresponds to the lowest-degree check node connected to exactly one punctured variable node. Dyn-PEBP uses empirically tuned $\gamma=0.35$; RD-RBP uses $\beta=0.95$ following \cite{Huilian2019Residual}.

\begin{figure}[H]
	\centering
	\includegraphics[width=0.96\columnwidth]{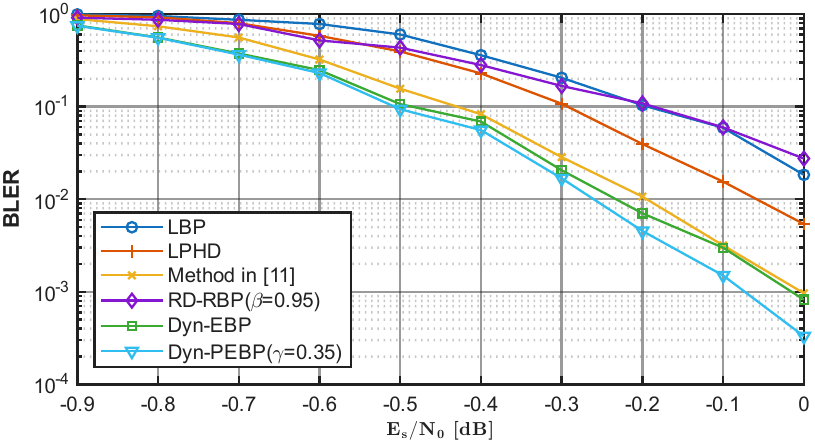}
	\caption{BLER comparison for the length-6144, rate-0.3438 LDPC code.}
	\label{fig:1482}
\end{figure}

\begin{figure}[H]
	\centering
	\includegraphics[width=0.96\columnwidth]{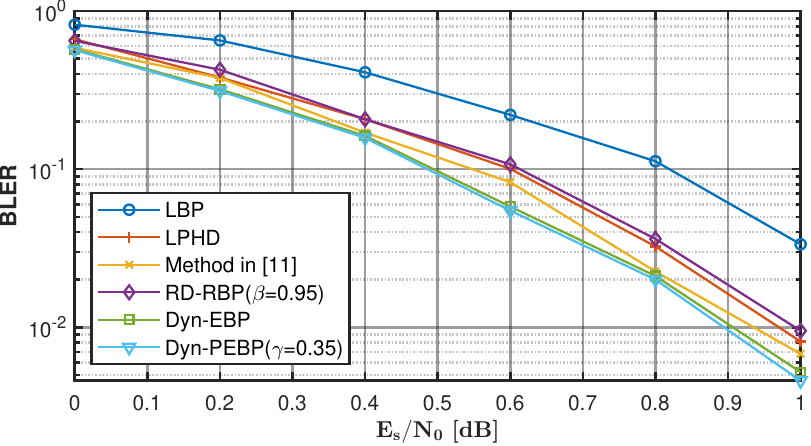}
	\caption{BLER comparison for the length-1482, rate-0.386 LDPC code.}
	\label{fig:1170}
\end{figure}

\begin{figure}[H]
	\centering
	\includegraphics[width=0.96\columnwidth]{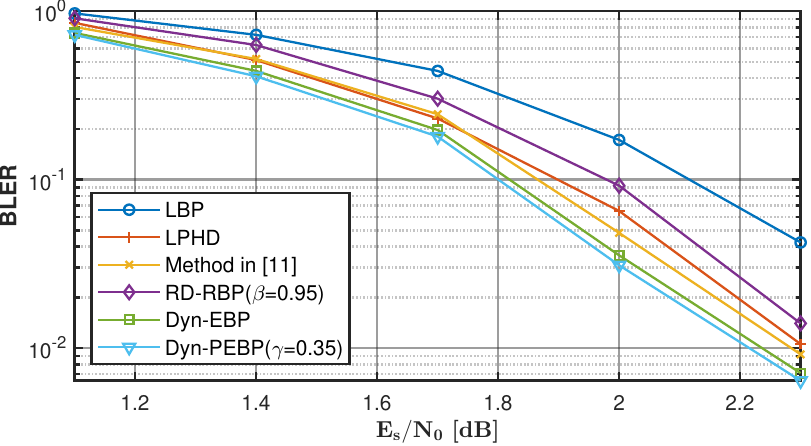}
	\caption{BLER comparison for the length-1170, rate-0.4681 LDPC code.}
	\label{fig:1320}
\end{figure}

\begin{figure}[H]
	\centering
	\includegraphics[width=0.96\columnwidth]{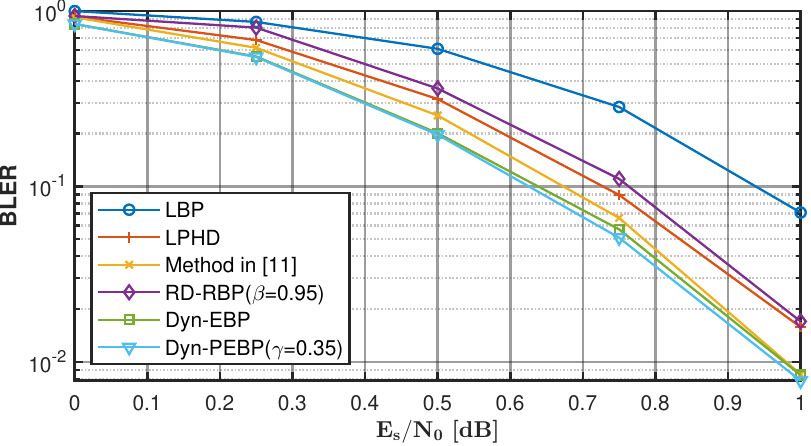}
	\caption{BLER comparison for the length-1320, rate-0.4 LDPC code.}
	\label{fig:2640}
\end{figure}

Figs.~\ref{fig:1482}--\ref{fig:2640} show that the proposed schedules remain competitive across the four settings. LBP is the weakest baseline, as expected from its fixed ordering. LPHD and the channel-only schedule in \cite{chang2025} capture part of the reliability structure induced by puncturing and degree distribution. Updating unsatisfied probabilities during message passing gives a slight reduction in BLER. The difference from \cite{chang2025} is thus an incremental refinement rather than a clear performance separation.

The RD-RBP comparison is informative. RD-RBP reacts to message residuals and targets locally active graph regions, whereas Dyn-EBP and Dyn-PEBP rank check nodes by the probability that their parity equations are already reliable. The resulting curves are close, which suggests that the main advantage of the proposed metric is not aggressive residual chasing but steadier use of reliable constraints. Dyn-PEBP is slightly better than Dyn-EBP because the penalty permits selected repeated updates when they are supported by the current probabilities, while still avoiding the repeated-update pathology illustrated in Fig.~\ref{exn}.

A more observation is that Dyn-PEBP shows clearer advantages in the low-rate and longer-blocklength regime. When the code rate is below $0.4$, the proposed schedule typically yields a stable gain of around $0.05$ dB at moderate BLER levels. As the code rate increases, the improvement becomes unstable or even negligible. This is mainly because higher-rate LDPC matrices tend to exhibit more regular structures, which reduces the diversity of node reliabilities and thereby weakens the effect of scheduling. For lower-rate and larger-blocklength cases, the graph exhibits stronger irregularity and more pronounced reliability heterogeneity, allowing dynamic scheduling to better exploit evolving LLR structures.

Across the four settings, the additional gain is most naturally explained by the evolution of the LLRs rather than by a different decoding rule. At low SNR, many parity checks have similar unsatisfied probabilities and the reliability-driven schedules tend to choose comparable early layers. As the SNR increases, the LLR magnitudes become more separated, and refreshing \eqref{cv} allows the decoder to distinguish layers that were close under the initial channel-only ordering. This produces small, repeatable BLER reductions without changing the underlying BP updates.

The penalty coefficient in Dyn-PEBP was kept fixed at $\gamma=0.35$ in all simulations. This choice avoids per-SNR tuning and gives a moderate compromise between the two extremes: a nearly greedy policy when $\gamma$ is very small and the coverage-oriented Dyn-EBP behavior when $\gamma$ is large. The results therefore indicate that a simple fixed penalty is sufficient for stable behavior over the tested blocklengths and code rates.

From an implementation standpoint, the proposed schedules do not alter the BP update equations or the stopping rule. Their additional cost is the local priority refresh after C2V updates, and the affected priorities are confined to check nodes adjacent to the updated variable nodes. This is particularly suitable for QC layer scheduling: the decoder can maintain one priority per layer by averaging the lifted check node probabilities, rather than managing all individual check nodes independently.

\section{Conclusions}\label{con}
This letter proposed Dyn-EBP and Dyn-PEBP, two dynamic LBP schedules that use check node unsatisfied probabilities as layer priorities. Dyn-EBP refreshes these probabilities during decoding while updating each check node once per iteration. Dyn-PEBP adds a linear update-count penalty, giving the decoder limited freedom to revisit useful checks without letting repeated updates dominate the schedule.

For 5G NR LDPC BG1 codes, the proposed schedules give small BLER gains relative to the channel-only probability schedule in \cite{chang2025} and remain competitive with LPHD and RD-RBP. The contribution should therefore be read as an ordering refinement: the BP equations are unchanged, and the improvement comes from making the probability criterion follow the current LLRs. This scope is useful in low-iteration decoding, where increasing the iteration count is often undesirable. Future work includes adaptive choices of $\gamma$, richer penalty functions, and parallel implementations.

\clearpage
\IEEEtriggeratref{9}

\end{document}